\documentclass[showpacs, showkeys,twocolumn,amsmath,amssymb,prb]{revtex4}

\usepackage{graphicx}
\usepackage{dcolumn}
\usepackage{bm}
\begin{document}
\newcommand{\ud}{d}
\newcommand{\hs}{\hspace*{0.5cm}}


\title{Heat capacity estimators for random series path-integral methods by finite-difference schemes}

\author{Cristian Predescu}
\author{Dubravko Sabo}
\author{J. D. Doll}
\affiliation{Department of Chemistry,  Brown University,  Providence,
Rhode Island 02912}
\author{David L. Freeman}
\affiliation{Department of Chemistry, University of Rhode Island,
Kingston, Rhode Island 02881}
\date{\today}

\begin{abstract} Previous heat capacity estimators used in path integral simulations  either have large variances that grow to infinity with the number of path variables or require the evaluation of first and second order derivatives of the potential. In the present paper, we show that the evaluation of the total energy by the T-method estimator and of the heat capacity by the TT-method estimator can be implemented by a finite difference scheme in a stable fashion. As such, the variances of the resulting estimators are finite and the evaluation of the estimators requires the potential function only. By comparison with the task of computing the partition function, the evaluation of the estimators requires $k + 1$ times more calls to the potential,  where $k$ is the order of the difference scheme employed. Quantum Monte Carlo simulations for the $\text{Ne}_{13}$ cluster demonstrate that a second order central-difference scheme should suffice for most applications. 
\end{abstract}


\maketitle

\section{Introduction} \label{sec:intro}

It is said that path integral methods transform a quantum equilibrium problem into a classical one by judicious use of dimensionality.\cite{Nig99} Yet, the computation of the average energy\cite{Bar79, Her82, Gia88, Cao89, Cep95, Fer95, Kol96, Jan97, Cha98, Dol99, Ele99, Gla02, Pre02} or the heat capacity\cite{Nei00a, Gla02a} of a quantum canonical ensemble reveals that the quantum-classical analogy is far from being trivial, even if distinguishable particles are assumed. One observes an increase in the computational time not only with the number of path variables considered, but also with the dimensionality of the system. This is so because estimators of finite variance usually involve  first or second order derivatives of the potential. The number of such derivatives scales linearly or quadratically with the number of degrees of freedom of the system. For example, numerical studies of even moderately large quantum clusters are severely hindered by this substantial increase in the number of quantities that must be evaluated.\cite{Nei00a} 

Recently, Predescu and Doll\cite{Pre02} have observed that a simple rescaling of the Brownian bridge entering the Feynman-Kac formula\cite{Fey48, Kac51, Sim79} produces path-integral techniques for which the dependence with the temperature of the path distributions is buried into the potential part of the imaginary-time action. Formal differentiation of the logarithm of the partition function leads to a special form of the thermodynamic estimator (T-method estimator) that does not have the variance  difficulties associated with  the Barker estimator for large numbers of path variables.\cite{Bar79, Her82} Even though the resulting T-method estimator closely resembles the virial estimators,\cite{Her82, Ele99, Gla02} it does not rely on the virial theorem to recover the kinetic energy. For instance, this T-method estimator produces correct results even for potentials that are not confined, e.g. a free particle. Therefore,  the variance of the T-method estimator is lower than that of the virial estimator because the classical part of the energy is explicitly introduced as a constant and is not obtained from the virial theorem. In a recent study of the $(\text{H}_2)_{22}$ cluster at the temperature of $6$~K,\cite{Pre03d} difficulties associated with the virial estimator for low-temperature systems\cite{Gia88, Fer95, Jan97} were not observed to appear for the T-method estimator introduced by Predescu and Doll. Such differences between the estimators are even more significant for the  heat capacity estimators and will be revealed in the present paper by comparing the statistical errors for the Predescu and Doll-type estimators with those for the double virial estimator.\cite{Nei00a} 

In order to avoid any confusion with earlier estimators, we mention that in the present article by  T-method and H-method estimators we understand the respective energy estimators introduced by Predescu and Doll in Ref.~\onlinecite{Pre02}. By TT-method and TH-method estimators, we understand the heat capacity estimators that are obtained from the corresponding energy estimators by temperature differentiation. 

The T-method estimator is closely related and similar in form to the centroid virial estimator.\cite{Cep95, Gla02, Gla02a} There are however two differences. First, the T-method estimator involves fluctuations of  the Brownian bridge relative to one arbitrary point. The centroid virial estimator involves similar fluctuations but relative to the path centroid. It can be shown that the ratio between the average square fluctuations of the Brownian bridge relative to some preferential point and to the path centroid is $2$.\cite{Pre02d} Thus, the two estimators have  similar behavior with the nature of the quantum system, the temperature, and the Monte Carlo sampling method, though the centroid virial estimator may exhibit a slightly lower variance.  

A second and more important difference, which constitutes the starting point of  the present development, is the fact that the T-method estimator is a veritable thermodynamic estimator, in the sense that it is obtained by temperature differentiation of the quantum partition function (however, as discussed in a previous paragraph, one needs to utilize a special form for the Feynman-Kac formula, with the temperature dependence of the path distribution buried into the potential).  The temperature differentiation can be  implemented numerically by a finite-difference scheme and leads to  numerically stable algorithms that do not require derivatives of the potential. This observation proves to be extremely important for  heat capacity calculations because formal temperature differentiation leads to expressions involving all first and second order derivatives of the potential. By numerical temperature differentiation, one obtains an important speed-up in the evaluation of the above mentioned thermodynamic properties, especially for large dimensional systems or for complicated potentials. 

In this article, we also propose an analytical heat capacity estimator  that involves the first derivatives of the potential only. This is obtained from the analytical form of the  TT-method estimator by an integration by parts suggested by Predescu and Doll in the derivation of their  special H-method energy estimator.\cite{Pre02} The two estimators, called in this paper the TT-method estimator and the modified TT-method estimator respectively, may have slightly different variances. As discussed in the previous paragraph, the first one is to be implemented by finite-difference, whereas for the second one we shall use exact analytical formulas. 

The relative merits of the new heat capacity estimators will be demonstrated for the $\text{Ne}_{13}$ cluster. For this example, we provide a comparison of the statistical errors of the new estimators with those of the double virial estimator utilized by Neirotti, Freeman, and Doll.\cite{Nei00a} We shall also  clarify a number of issues raised in the Neirotti, Freeman, and Doll study of this neon cluster. The numerical simulation presented serves to demonstrate the power of the path integral approach utilized as well as to provide essentially exact numerical data necessary for comparison in the development of quantum approximations that can be employed for larger or more complicated systems.\cite{Cal01}

\section{Thermodynamic energy and heat capacity estimators}

In this section, we derive the formal expressions for the heat capacity of a $d$-dimensional canonical quantum mechanical system made up of distinguishable particles. The particles have masses $\{m_{0,i}; \; 1\leq i \leq d\}$ and move in the potential $V(\mathbf{x})$. The vector $\mathbf{x}$, the transpose of which is $\mathbf{x}^{\text{T}}=  (x_1, \ldots, x_d)$, denotes the position of the particles in the configuration space $\mathbb{R}^d$. The canonical system is characterized by inverse temperature $\beta = 1/ (k_{B} T)$. Its average energy and heat capacity can be obtained by temperature differentiation of the partition function $Z(\beta)$, producing the formulas
\begin{equation}
\label{eq:II1}
\left\langle E \right\rangle_\beta^T = - \frac{1}{Z(\beta)} \frac{dZ(\beta)}{d\beta}
\end{equation}
and
\begin{equation}
\label{eq:II2}
\left\langle C_V \right\rangle_\beta^{TT} = k_B  \left\{ \frac{\beta^2}{Z(\beta)} \frac{d^2Z(\beta)}{d\beta^2}-\left[\frac{\beta}{Z(\beta)} \frac{dZ(\beta)}{d\beta}\right]^2\right\},
\end{equation}
respectively. The partition function of the system  is obtained as the integral over the configuration space of the diagonal density matrix
\begin{equation}
\label{eq:II3}
Z(\beta) = \int_{\mathbb{R}^d} \rho(\mathbf{x};\beta)d\mathbf{x}. 
\end{equation}

In the path-integral approach, the density matrix is evaluated with the help of the Feynman-Kac formula. We split the present section into two parts. In the first part, we shall discuss the random series implementation of the Feynman-Kac formula and introduce some relevant notation. In the second part, we deduce the formal expression of the TT-method heat capacity estimator and discuss its numerical implementation by finite-difference schemes. Then, we derive the modified TT-method estimator, the analytical expression of which involves first-order derivatives of the potential only.

\subsection{Random series path integral techniques}
In the random series implementation of the Feynman-Kac formula, the  density matrix of a one-dimensional quantum system is obtained as follows.\cite{Pre02}  Let $\{\lambda_k(\tau)\}_{k \geq 1}$ be a system of functions on the interval $[0,1]$ that, together with the constant
function $\lambda_0(\tau)=1$, make up an orthonormal basis in $L^2[0,1]$.
Define
\[ \Lambda_k(t)=\int_0^t \lambda_k(u)\ud u.\] Let $\Omega$ denote the space of
infinite sequences $\bar{a}\equiv(a_1,a_2,\ldots)$ and let
\begin{equation}
\label{eq:II4}
\ud P[\bar{a}]=\prod_{k=1}^{\infty}\ud \mu(a_k)
\end{equation}
be the probability measure on $\Omega$ such that the coordinate maps
$\bar{a}\rightarrow a_k$ are independent identically distributed (i.i.d.)
variables with distribution probability
\begin{equation}
\label{eq:II5}
\ud \mu(a_i)= \frac{1}{\sqrt{2\pi}} e^{-a_i^2/2}\,\ud a_i.
\end{equation} 
Then,  the Feynman-Kac formula reads\cite{Pre02} 
\begin{eqnarray}
\label{eq:II6}
    \frac{\rho(x, x' ;\beta)}{\rho_{fp}(x, x'  ;\beta)}&=&\int_{\Omega}\ud
P[\bar{a}]\nonumber  \exp\bigg\{-\beta
\int_{0}^{1}\! \!  V\Big[x_r(u) \\& +& \sigma \sum_{k=1}^{\infty}a_k
\Lambda_k(u) \Big]\ud u\bigg\},
\end{eqnarray} 
where $x_r(u) = x + (x'-x)u$ and $\sigma = (\hbar^2 \beta/m_0)^{1/2}$. The quantity $\rho_{fp}(x, x'  ;\beta)$ represents the density matrix for a similar free particle. The series 
\[
B_u^0(\bar{a})= \sum_{k=1}^{\infty}a_k \Lambda_k(u)
\]
represents a stochastic process equal in distribution to a standard Brownian bridge. 

 For a $d$-dimensional system, the Feynman-Kac formula
is obtained by employing an independent random series for each additional
degree of freedom. As such, we consider the space $\Omega^d$ made up of all sequences  $\bar{\mathbf{a}}\equiv(\mathbf{a}_1,\mathbf{a}_2,\ldots)$ of vectors 
\[\mathbf{a}_k = \left(\begin{array}{c} a_{1,k} \\ \vdots\\ a_{d,k}\end{array}\right)\]
and denote the line $i$ of $\bar{\mathbf{a}}$ by  $\bar{a}_i = (a_{i,0}, a_{i,1}, \ldots)$. On the space $\Omega^d$, we define the probability measure
\begin{equation}
\label{eq:II7}
\ud P[\bar{\mathbf{a}}]=\prod_{i=1}^{d}\ud P[\bar{a}_i] 
\end{equation}
with
\[
\ud P[\bar{a}_i] = \prod_{k=1}^{\infty}\ud \mu(a_{i,k}).
\]
Under this probability measure, the coordinate maps
$\bar{\mathbf{a}}\rightarrow a_{i,k}$ are i.i.d. standard normal variables. We also consider the vector $\mathbf{\sigma}^\text{T} = (\sigma_1, \ldots, \sigma_d)$ of components $\sigma_i = (\hbar^2\beta / m_{0,i})^{1/2}$ and let $\mathbf{x}_r(u) = \mathbf{x}+(\mathbf{x}'-\mathbf{x})u$ be a straight line connecting the points $\mathbf{x}$ and $\mathbf{x}'$.
Then,  the Feynman-Kac formula reads 
\begin{eqnarray}
\label{eq:II8}
    \frac{\rho(\mathbf{x}, \mathbf{x}' ;\beta)}{\rho_{fp}(\mathbf{x}, \mathbf{x}'  ;\beta)}&=&\int_{\Omega^d}\ud
P[\bar{\mathbf{a}}]\nonumber  \exp\bigg\{-\beta
\int_{0}^{1}\! \!  V\Big[\mathbf{x}_r(u) \\& +& \sigma \sum_{k=1}^{\infty}\mathbf{a}_k
\Lambda_k(u) \Big]\ud u\bigg\},
\end{eqnarray} 
where 
\[\sigma \mathbf{a}_k = \left(\begin{array}{c}\sigma_1 a_{1,k}\\ \vdots \\ \sigma_d a_{d,k}\end{array}\right).\]
The series
\[
B_u^0(\bar{\mathbf{a}})= \sum_{k=1}^{\infty}\mathbf{a}_k \Lambda_k(u)
\]
is equal in distribution to a  $d$-dimensional Brownian bridge (a vector-valued stochastic process whose components are independent one-dimensional Brownian bridges).

In practical applications, one cannot work with the full random series implementation of the Feynman-Kac formula. Instead, one considers finite-dimensional approximations to Eq.~(\ref{eq:II8}), the simplest of which have the general form
\begin{eqnarray}
\label{eq:II9}&&
\frac{\rho_n(\mathbf{x}, \mathbf{x}' ;\beta)}{\rho_{fp}(\mathbf{x}, \mathbf{x}'
;\beta)}=\int_{\Omega^d} dP[\bar{\mathbf{a}}]\nonumber  \\&& \times \exp\bigg\{-\beta \; \int_{0}^{1}\! \!
V\Big[\mathbf{x}_r(u)+ \sigma \sum_{k=1}^{qn+p}\mathbf{a}_k
\tilde{\Lambda}_{n,k}(u)\Big]\ud u\bigg\},\qquad
\end{eqnarray} where $q$ and $p$ are some fixed integers. The functions $\tilde{\Lambda}_{n,k}$ are chosen so that to maximize  the rate of convergence
\[
\rho_n(\mathbf{x}, \mathbf{x}' ;\beta) \to \rho(\mathbf{x}, \mathbf{x}' ;\beta).
\]
Though the problem of maximizing the order of convergence is still far from a final resolution, several schemes in the larger class of reweighted techniques were proven to have cubic or quartic asymptotic convergence.\cite{Pre03, Pre03e} The construction of the functions $\tilde{\Lambda}_{n,k}$ and of associated quadrature schemes for the computation of the path averages appearing in Eq.~(\ref{eq:II9}) have been discussed elsewhere.\cite{Pre03, Pre03e, Pre03b} For the numerical examples presented in this article, we  use a so-called L\'evy-Ciesielski reweighted path integral method having quartic convergence.\cite{Pre03e}  To a large extent, the analytical expressions of the functions $\tilde{\Lambda}_{n,k}(u)$ and the nature of the quadrature schemes are not important for the present development. For more information, the reader is advised to consult the cited references.

To simplify notation, we introduce several auxiliary quantities $B_{u,n}^0(\bar{\mathbf{a}})$, $U_n(\mathbf{x},\mathbf{x}',\bar{\mathbf{a}};\beta)$, and  $X_n(\mathbf{x},\mathbf{x}',\bar{\mathbf{a}};\beta)$,
defined by the expressions
\begin{equation}
\label{eq:II10}
B_{u,n}^0(\bar{\mathbf{a}})= \sum_{k=1}^{qn+p}\mathbf{a}_k
\tilde{\Lambda}_{n,k}(u),
\end{equation}
\begin{equation}
\label{eq:II11} U_n(\mathbf{x},\mathbf{x}',\bar{\mathbf{a}};\beta)=\int_{0}^{1}\! \! V\Big[\mathbf{x}_r(u)+ \sigma
B_{u,n}^0(\bar{\mathbf{a}})\Big]\ud u,
\end{equation} and
\begin{equation}
\label{eq:II12} 
X_n(\mathbf{x},\mathbf{x}',\bar{\mathbf{a}};\beta)=\rho_{fp}(\mathbf{x},\mathbf{x}';\beta)
 \exp\left[-\beta U_n(\mathbf{x},\mathbf{x}',\bar{\mathbf{a}};\beta)\right],
\end{equation} respectively. The similar relations for the full Feynman-Kac formula are denoted by $B_{u}^0(\bar{\mathbf{a}})$, $U_\infty(\mathbf{x},\mathbf{x}',\bar{\mathbf{a}};\beta)$, and  $X_\infty(\mathbf{x},\mathbf{x}',\bar{\mathbf{a}};\beta)$, respectively.
With the new notation, Eq.~(\ref{eq:II9})
becomes
\begin{eqnarray}
\label{eq:II13}&&
\rho_n(\mathbf{x}, \mathbf{x}' ;\beta)=\int_{\Omega^d}\ud
P[\bar{\mathbf{a}}]X_n(\mathbf{x},\mathbf{x}',\bar{\mathbf{a}};\beta),
\end{eqnarray} 
whereas the Feynman-Kac formula reads
\begin{eqnarray}
\label{eq:II14}&&
\rho(\mathbf{x}, \mathbf{x}' ;\beta)=\int_{\Omega^d}\ud
P[\bar{\mathbf{a}}]X_\infty(\mathbf{x},\mathbf{x}',\bar{\mathbf{a}};\beta).
\end{eqnarray}

In this paper, we make the convention that $\mathbf{x}'$ is dropped whenever $\mathbf{x} = \mathbf{x}'$. In order to arrive at the definition of the energy and the heat capacity estimators, 
it is convenient to introduce the quantities
\begin{eqnarray}
\label{eq:II15} &&
\nonumber  R_n(\mathbf{x},\bar{\mathbf{a}};\beta,\epsilon)= \frac{X_n(\mathbf{x},\bar{\mathbf{a}};\beta \epsilon)}{X_n(\mathbf{x},\bar{\mathbf{a}};\beta)} = 
\epsilon^{-d/2}\\ && \times \exp\left[-\beta\epsilon U_n(\mathbf{x},\bar{\mathbf{a}};\beta\epsilon)  + \beta U_n(\mathbf{x},\bar{\mathbf{a}};\beta)\right]
\end{eqnarray} and
\begin{eqnarray}
\label{eq:II16} &&
\nonumber  R_\infty(\mathbf{x},\bar{\mathbf{a}};\beta,\epsilon)= \frac{X_\infty(\mathbf{x},\bar{\mathbf{a}};\beta \epsilon)}{X_\infty(\mathbf{x},\bar{\mathbf{a}};\beta)} = 
\epsilon^{-d/2}\\ && \times \exp\left[-\beta\epsilon U_\infty(\mathbf{x},\bar{\mathbf{a}};\beta\epsilon)  + \beta U_\infty(\mathbf{x},\bar{\mathbf{a}};\beta)\right], \qquad
\end{eqnarray} respectively. 
We have
\begin{widetext}
\begin{eqnarray}
\label{eq:II17}
\frac{\beta}{Z(\beta)}\frac{d Z(\beta)}{d\beta} = \frac{\int_{\mathbb{R}^d}\ud \mathbf{x}\int_{\Omega^d}\ud
P[\bar{\mathbf{a}}]X_\infty(\mathbf{x},\bar{\mathbf{a}};\beta) \frac{d}{d\epsilon} R_\infty(\mathbf{x},\bar{\mathbf{a}};\beta,\epsilon)\Big|_{\epsilon = 1} }{\int_{\mathbb{R}^d}\ud \mathbf{x}\int_{\Omega^d}\ud
P[\bar{\mathbf{a}}]X_\infty(\mathbf{x},\bar{\mathbf{a}};\beta)}
\end{eqnarray}
and 
\begin{eqnarray}
\label{eq:II18}
\frac{\beta^2}{Z(\beta)}\frac{d^2 Z(\beta)}{d\beta^2} =  \frac{\int_{\mathbb{R}^d}\ud \mathbf{x}\int_{\Omega^d}\ud
P[\bar{\mathbf{a}}]X_\infty(\mathbf{x},\bar{\mathbf{a}};\beta) \frac{d^2}{d\epsilon^2} R_\infty(\mathbf{x},\bar{\mathbf{a}};\beta,\epsilon)\Big|_{\epsilon = 1} }{\int_{\mathbb{R}^d}\ud \mathbf{x}\int_{\Omega^d}\ud
P[\bar{\mathbf{a}}]X_\infty(\mathbf{x},\bar{\mathbf{a}};\beta)}.
\end{eqnarray}
\end{widetext}
The quantities above can be evaluated by Monte Carlo integration as the limit $n \to \infty$ of the sequences
\begin{widetext}
\begin{eqnarray}
\label{eq:II19}
\frac{\beta}{Z_n(\beta)}\frac{d Z_n(\beta)}{d\beta} =   \frac{\int_{\mathbb{R}^d}\ud \mathbf{x}\int_{\Omega^d}\ud
P[\bar{\mathbf{a}}]X_n(\mathbf{x},\bar{\mathbf{a}};\beta) \frac{d}{d\epsilon} R_n(\mathbf{x},\bar{\mathbf{a}};\beta,\epsilon)\Big|_{\epsilon = 1} }{\int_{\mathbb{R}^d}\ud \mathbf{x}\int_{\Omega^d}\ud
P[\bar{\mathbf{a}}]X_n(\mathbf{x},\bar{\mathbf{a}};\beta)}
\end{eqnarray}
and 
\begin{eqnarray}
\label{eq:II20}
\frac{\beta^2}{Z_n(\beta)}\frac{d^2 Z_n(\beta)}{d\beta^2} =  \frac{\int_{\mathbb{R}^d}\ud \mathbf{x}\int_{\Omega^d}\ud
P[\bar{\mathbf{a}}]X_n(\mathbf{x},\bar{\mathbf{a}};\beta) \frac{d^2}{d\epsilon^2} R_n(\mathbf{x},\bar{\mathbf{a}};\beta,\epsilon)\Big|_{\epsilon = 1} }{\int_{\mathbb{R}^d}\ud \mathbf{x}\int_{\Omega^d}\ud
P[\bar{\mathbf{a}}]X_n(\mathbf{x},\bar{\mathbf{a}};\beta)},
\end{eqnarray}
\end{widetext}
respectively.

In the finite-difference scheme that is advocated in this paper, the derivatives against $\epsilon$ appearing in Eqs.~(\ref{eq:II19}) and (\ref{eq:II20}) may be evaluated numerically by  finite difference. Such an approach is expected to be much faster than the analytical evaluation of the derivatives, especially for large dimensional systems or systems with complicated potentials. Though higher order central-difference schemes can be employed, a second order scheme produces 
\begin{eqnarray*}&&\nonumber
\frac{d}{d\epsilon} R_n(\mathbf{x},\bar{\mathbf{a}};\beta,\epsilon)\Big|_{\epsilon = 1}  \approx (2\epsilon_0)^{-1} \\ && \times \left[R_n(\mathbf{x},\bar{\mathbf{a}};\beta,1+\epsilon_0)- R_n(\mathbf{x},\bar{\mathbf{a}};\beta,1-\epsilon_0)\right] 
\end{eqnarray*}
and
\begin{eqnarray*} \nonumber
\frac{d^2}{d\epsilon^2} R_n(\mathbf{x},\bar{\mathbf{a}};\beta,\epsilon)\Big|_{\epsilon = 1} \approx \epsilon_0^{-2} \left[R_n(\mathbf{x},\bar{\mathbf{a}};\beta,1+\epsilon_0)\right. \\  
- \left. 2 R_n(\mathbf{x},\bar{\mathbf{a}};\beta,1) + R_n(\mathbf{x},\bar{\mathbf{a}};\beta,1-\epsilon_0)\right],
\end{eqnarray*}
with  error of  order $O(\epsilon_0^2)$. Such a direct approach may prove useful for highly quantum systems or for pathological systems, as for instance a particle in a box. However, for smooth enough potentials, the alternatives that are analyzed in the following subsection may prove to be superior. 

\subsection{Expressions of the heat capacity estimators}

In this subsection, we shall put the relevant quantities entering the expression of the heat capacity estimator in a form that is exact in the high-temperature limit or in the limit that the physical system is classical. For this purpose, we assume that $\exp[-\beta V(\mathbf{x})]$ has second order Sobolev derivatives as a function of $\mathbf{x}$. In the second part of the present subsection, we shall derive a special analytical expression for the heat capacity estimator that employs the first order derivatives of the potential, only. This modified heat capacity estimator gives results identical to the first one, but it may have a slightly different variance. 

By explicit computation, one argues that
\begin{eqnarray}
\label{eq:II21}&&\nonumber
\frac{d}{d\epsilon} R_n(\mathbf{x},\bar{\mathbf{a}};\beta,\epsilon)\Big|_{\epsilon = 1} \\ && =  -\frac{d}{2} - \beta U_n(\mathbf{x},\bar{\mathbf{a}};\beta)- \beta \frac{d}{d\epsilon} U_n(\mathbf{x},\bar{\mathbf{a}};\beta\epsilon)\Big|_{\epsilon = 1} \qquad
\end{eqnarray}
and
\begin{eqnarray}
\label{eq:II22}&&\nonumber
\frac{d^2}{d\epsilon^2} R_n(\mathbf{x},\bar{\mathbf{a}};\beta,\epsilon)\Big|_{\epsilon = 1}  = \left[\frac{d}{d\epsilon} R_n(\mathbf{x},\bar{\mathbf{a}};\beta,\epsilon)\Big|_{\epsilon = 1}\right]^2 \\ && + \frac{d}{2} - 2\beta \frac{d}{d\epsilon} U_n(\mathbf{x},\bar{\mathbf{a}};\beta\epsilon)\Big|_{\epsilon = 1}- \beta \frac{d^2}{d\epsilon^2} U_n(\mathbf{x},\bar{\mathbf{a}};\beta\epsilon)\Big|_{\epsilon = 1}. \qquad
\end{eqnarray}
The first and second derivatives of the function $U_n(\mathbf{x},\bar{\mathbf{a}};\beta\epsilon)$ around the point $\epsilon = 1$ can be evaluated by finite difference, as shown in the preceding subsection. However, we notice that in the limit that the physical system behaves classically, $U_n(\mathbf{x},\bar{\mathbf{a}};\beta\epsilon) \approx V(\mathbf{x})$ and the derivatives against $\epsilon$ vanish. Moreover, in this limit any finite-difference scheme produces the exact classical results. It is therefore apparent that the utilization of the derivatives of the functions $U_n(\mathbf{x},\bar{\mathbf{a}};\beta\epsilon)$ instead of the derivatives of $R_n(\mathbf{x},\bar{\mathbf{a}};\beta,\epsilon)$ has certain numerical advantages, increasing the range of acceptable values for the discretization step $\epsilon_0$.

We now proceed and compute the analytical expression of the derivatives of the function $U_n(\mathbf{x},\bar{\mathbf{a}};\beta\epsilon)$. We have
\begin{eqnarray}
\label{eq:II23}&&\nonumber
 \frac{d}{d\epsilon} U_n(\mathbf{x},\bar{\mathbf{a}};\beta\epsilon)\Big|_{\epsilon = 1}= \\ &&
\frac{1}{2} \sum_{i=1}^d \sigma_i \int_{0}^{1}\! \! \partial_iV\Big[\mathbf{x}+ \sigma
B_{u,n}^0(\bar{\mathbf{a}})\Big]B_{u,n}^{0,i}(\bar{\mathbf{a}})\ud u,
\end{eqnarray}
where 
\[
B_{u,n}^{0,i}(\bar{\mathbf{a}}) = \sum_{k=1}^{qn+p}a_{i,k}
\tilde{\Lambda}_{n,k}(u).
\]
One also computes
\begin{eqnarray}
\label{eq:II24}&&\nonumber
\frac{d^2}{d\epsilon^2} U_n(\mathbf{x},\bar{\mathbf{a}};\beta\epsilon)\Big|_{\epsilon = 1} = -\frac{1}{4} \sum_{i=1}^d \sigma_i \\ && \nonumber \times \int_{0}^{1}\! \! \partial_iV\Big[\mathbf{x}+ \sigma
B_{u,n}^0(\bar{\mathbf{a}})\Big]B_{u,n}^{0,i}(\bar{\mathbf{a}})\ud u  + \frac{1}{4} \sum_{i,j =1}^d \sigma_i \sigma_j \\ && \times \int_{0}^{1}\! \! \partial_{i,j}V\Big[\mathbf{x}+ \sigma
B_{u,n}^0(\bar{\mathbf{a}})\Big]B_{u,n}^{0,i}(\bar{\mathbf{a}})B_{u,n}^{0,j}(\bar{\mathbf{a}})\ud u, \qquad
\end{eqnarray}

The expression given by Eq.~(\ref{eq:II24}) is not computationally very convenient because it requires the evaluation of $d(d+1)/2$ path averages for as many different second-order derivatives 
\[
\int_{0}^{1}\! \! \partial_{i,j}V\Big[\mathbf{x}+ \sigma
B_{u,n}^0(\bar{\mathbf{a}})\Big]B_{u,n}^{0,i}(\bar{\mathbf{a}})B_{u,n}^{0,j}(\bar{\mathbf{a}})\ud u.
\]
It is for this reason that we advocate the use of a finite difference scheme instead of the analytical formulas. For large enough physical systems or for complicated potentials for which the derivatives are not readily available, the finite difference scheme will enjoy a significant computational advantage. Parenthetically, Eq.~(\ref{eq:II24}) shows that the TT-method heat capacity estimator is similar in form to the double virial heat capacity estimator\cite{Nei00a} or to the centroid double virial heat capacity estimator.\cite{Gla02a} However, it has the distinctive feature (characteristic of the Barker estimators) that it can be implemented by a finite-difference scheme, yet it maintains to a good degree the low variance of the centroid double virial estimator.

For strongly quantum systems, as for instance low temperature hydrogen or helium clusters, there is sometimes the need to validate the convergence of the path integral methods by employing the agreement between the T-method and the H-method energy estimators.\cite{Pre03d} As shown by Predescu and Doll,\cite{Pre02} the H-method estimator can be put into the ``force-force correlation'' form by a simple integration by parts. This form requires the first order derivatives of the potential only. In such cases, given that the first order derivatives of the potential are computed anyway, it would be advantageous if we could evaluate the heat capacity as a functional of these derivatives only. This can actually be done (again by integration by parts) as follows. Observe that
\begin{widetext}
\begin{eqnarray}
\label{eq:II25}&&\nonumber
\int_{\mathbb{R}}dx_j e^{-\beta \int_{0}^{1}\! \! V\left[\mathbf{x}+ \sigma
B_{u,n}^0(\bar{\mathbf{a}})\right]\ud u}\int_{0}^{1}\! \! \partial_{i,j}V\Big[\mathbf{x}+ \sigma
B_{u,n}^0(\bar{\mathbf{a}})\Big]B_{u,n}^{0,i}(\bar{\mathbf{a}})B_{u,n}^{0,j}(\bar{\mathbf{a}})\ud u \\ && = 
\int_{\mathbb{R}}dx_j e^{-\beta \int_{0}^{1}\! \! V\left[\mathbf{x}+ \sigma
B_{u,n}^0(\bar{\mathbf{a}})\right]\ud u} \beta \left\{\int_{0}^{1}\! \!  \partial_{i}V\Big[\mathbf{x}+ \sigma
B_{u,n}^0(\bar{\mathbf{a}})\Big]B_{u,n}^{0,i}(\bar{\mathbf{a}})B_{u,n}^{0,j}(\bar{\mathbf{a}})\ud u\right\}\left\{\int_{0}^{1}\! \!  \partial_{j}V\Big[\mathbf{x}+ \sigma
B_{u,n}^0(\bar{\mathbf{a}})\Big]\ud u\right\}.
\end{eqnarray}
\end{widetext} 
Therefore, Eq.~(\ref{eq:II24}) can be replaced for the purpose of evaluating the heat capacity by 
\begin{eqnarray}
\label{eq:II26}&&\nonumber
\frac{d^2}{d\epsilon^2} U_n(\mathbf{x},\bar{\mathbf{a}};\beta\epsilon)\Big|_{\epsilon = 1} \equiv -\frac{1}{4} \sum_{i=1}^d \sigma_i \\ && \nonumber \times \int_{0}^{1}\! \! \partial_iV\Big[\mathbf{x}+ \sigma
B_{u,n}^0(\bar{\mathbf{a}})\Big]B_{u,n}^{0,i}(\bar{\mathbf{a}})\ud u  + \frac{\beta}{4} \sum_{i,j =1}^d \sigma_i \sigma_j \\ && \times \left\{\int_{0}^{1}\! \!  \partial_{i}V\Big[\mathbf{x}+ \sigma
B_{u,n}^0(\bar{\mathbf{a}})\Big]B_{u,n}^{0,i}(\bar{\mathbf{a}})B_{u,n}^{0,j}(\bar{\mathbf{a}})\ud u\right\}\\&& \nonumber
\times\left\{\int_{0}^{1}\! \!  \partial_{j}V\Big[\mathbf{x}+ \sigma
B_{u,n}^0(\bar{\mathbf{a}})\Big]\ud u\right\}. \qquad
\end{eqnarray}
We utilize the sign of equivalence $\equiv$ in the relation above to warn the reader that the equality implied by Eq.~(\ref{eq:II26}) does \emph{not} hold in the strict sense. Rather, it means that the expression to the right of the sign $\equiv$ produces estimates identical to the ones obtained by employing Eq.~(\ref{eq:II24}), though it may exhibit a different variance. The resulting heat capacity estimator will be called the modified TT-method estimator and will be denoted by $\left\langle C_V \right\rangle_\beta^{mTT} $ henceforth. Eq.~(\ref{eq:II26}) still involves $d^2$ path averages to be computed (which may become prohibitive for large dimensional systems), but this time the averages involve quantities that are computed anyway. Expensive calls to the second order derivatives of the potential are avoided.

\section{A numerical example}

We shall test the merits of the two heat capacity estimators discussed in the previous paragraph on a cluster of $N_p = 13$ neon atoms using a special path integral technique introduced in  Ref.~\onlinecite{Pre03e} and having quartic asymptotic convergence with respect to the number of path variables. The numerical implementation of this method is similar to the L\'evy-Ciesielski reweighted method utilized in Ref.~\onlinecite{Pre03d} and will not be reviewed here. 

Quantum studies of small Lennard-Jones neon clusters ($N_p \leq 7$) by ground-state\cite{Lei91, Ric91, Ric91a} or finite-temperature methods\cite{Bec90, Fra92} have revealed that the quantum effects are important, leading to large zero-point energies. By comparison, studies of larger clusters are relatively scant. The $\text{Ne}_{13}$ cluster is interesting because it is the smallest Lennard-Jones cluster that presents an effective classical melting point (at about $10$~K) marking a transition from a rigid to a liquid-like phase. The pronounced quantum effects have been found to lower the transition temperature by about $1$~K.\cite{Cha94, Nei00a} However, quantum heat capacities reported in literature and  computed   by path integral methods\cite{Cha94, Nei00a} or semiclassical techniques\cite{Cal01} are  not sufficiently accurate due to large statistical or systematic errors. To demonstrate the advantage of the new estimators, we propose to compute the heat capacity of the $\text{Ne}_{13}$ cluster over the range of temperatures $4-14$~K, with a statistical error (defined in the present article as two times the standard deviation) smaller than $1k_B$.  Such relatively accurate data are necessary for the development of approximate methods that can be employed for larger or more complicated systems.\cite{Cal01} They also constitute a realistic testbed for present and future path integral techniques. For comparison purposes, the best known data computed by the double virial estimator have a statistical error of about $10 k_B$ in the low temperature region.\cite{Nei00a} 

The total potential energy of the $\text{Ne}_{13}$ cluster is given by
\begin{equation}
\label{2.1} V_{tot} = \sum_{i<j}^{N_p} V_{LJ}(r_{ij})+\sum_{i=1}^{N_p}
V_c(\mathbf{r_i}),
\end{equation} where $V_{LJ}(r_{ij})$ is the pair interaction of
Lennard-Jones potential
\begin{equation}
\label{2.2} V_{LJ}(r_{ij}) = 4\epsilon_{LJ}\left
        [\left( \frac{\sigma_{LJ}}{r_{ij}}\right)^{12}
       -\left( \frac{\sigma_{LJ}}{r_{ij}}\right)^{6}\right]
\end{equation} and V$_{c}(\mathbf{r_i})$ is the confining potential
\begin{equation}
\label{2.3}
V_c(\mathbf{r_i})=\epsilon_{LJ}\left(\frac{|\mathbf{r_i}-\mathbf{R_{cm}}|}{R_c}\right)^{20}.
\end{equation}
   The values of the Lennard-Jones parameters
$\sigma_{LJ}$ and $\epsilon_{LJ}$ used are 2.749 {\AA} and 35.6 K,
respectively.\cite{Nei00a} The mass of the Ne atom was set to $m_0=20.0$, the rounded atomic mass of the most abundant isotope.
$\mathbf{R_{cm}}$ is the coordinate  of the center of mass of the
cluster  and is given by
\begin{equation}
\label{2.4}
\mathbf{R_{cm}}=\frac{1}{N_p}\sum_{i=1}^{N_p} \mathbf{r_i}.
\end{equation} Finally, $R_c=2\sigma_{LJ}$ is the confining radius.
The role of the confining potential $V_c(\mathbf{r_i})$ is to prevent
atoms from permanently leaving the cluster since the cluster in
vacuum at any finite temperature is metastable with respect to
evaporation.

The optimal choice of the parameter $R_c$ for the constraining potential
has been discussed in recent work.\cite{Nei00}  If $R_c$ is taken to be
too small, the properties of the system become sensitive to its choice,
whereas large values of $R_c$ can result in problems attaining an ergodic
simulation.  To facilitate comparisons, in the current work, $R_c$ has
been chosen to be identical to that used in Ref.~\onlinecite{Nei00a}.

\subsection{Sampling strategy}

The sampling strategy utilized in the present paper is similar to the one employed in Ref.~\onlinecite{Pre03d} except for the use of parallel tempering\cite{Mar92, Gey95, Tes96, Han97, Wu99, Fal99, Yan99} to cope with possible ergodicity problems. We have utilized a number of $42$ parallel streams, each running a replica of the system at a different temperature. For each stream, the basic Monte Carlo steps  consist in  moves of the physical coordinate $\textbf{x}_i$ of an individual particle together with the first one quarter of the associated path variables  or of the last three quarters of the path variables for the respective particle.  Eq.~(27) of Ref.~\onlinecite{Pre03e}, as specialized for the short-time approximation constructed in Section~IV.B of the same reference, shows that the first one quarter of the path variables are associated with Schauder functions, whereas the last three quarters are special functions  designed to maximize the asymptotic rate of convergence of the path integral method employed. Given the analytical differences between the Schauder and the special functions, one expects that the optimal maximal displacements for the path variables associated with functions from the two categories are  different. We mention that a poor sampling of the path variables associated with the special functions might ruin the quartic asymptotic convergence of the path integral method employed. For this reason, we attempt to update the path variables associated with the Schauder functions and with the special functions separately. The physical coordinate $\mathbf{x}_i$ is updated together with the Schauder functions.  Distinct acceptance ratios are computed for the two steps. The maximal displacements are adjusted in the equilibration phase of the computation so that each of the acceptance ratios eventually lies between $40\%$ and $60\%$. 

 The basic computational unit is the \emph{pass}, defined as the minimal set of Monte Carlo attempts over all variables in the system. Thus, a pass consists of $2 \cdot 13 = 26$ basic steps. Each Monte Carlo attempt is accepted or rejected according to the Metropolis logic.\cite{Met53, Kal86}  One defines a \emph{block} as a computational unit made up of forty thousand passes. The length of the blocks is large enough to ensure independence between the block-averages of various quantities computed. This independence has been checked with statistical tests, as described in Ref.~\onlinecite{Pre03d}.
As opposed to the computation performed in Ref.~\onlinecite{Pre03d}, the correlation between block averages of different streams has not been tested for independence. The explanation is that these block averages are correlated by the parallel tempering algorithm. However, we have tried to minimize this correlation by employing separate random number generators for each streams. These random number generators are obtained with the help of the Dynamic Creator package\cite{Mat98a, Mat98b} and produce highly independent streams of random numbers, as demonstrated by the statistical tests performed in Ref.~\onlinecite{Pre03d}. 

A swap between streams of neighboring temperatures has been attempted every $25$ passes and it has been accepted or rejected according to the parallel tempering logic.\cite{Mar92, Gey95, Tes96, Han97, Wu99, Fal99, Yan99} Any given stream attempts a swap with the neighboring streams of lower and higher temperatures in succession.  Because of this swapping strategy,  the streams of minimum and maximum temperatures are involved in swaps every $50$ steps, only. The interval $[4,14]$ has been divided in $40$ equal subintervals demarked by $41$ intermediate temperatures. Thus, the lowest temperature stream has run at a temperature of $4- (14 - 4)/40 = 3.75$~K. The efficiency of the parallel tempering algorithm depends strongly on how much the distributions for neighboring temperatures overlap. In classical simulations, the width of the overlap is inversely proportional to  the difference between inverse neighboring temperatures. It appears then that the optimal division of the interval $[4,14]$ involves equally spaced inverse temperature subintervals.  While not the optimal one, our choice of equal temperature subintervals has the advantage that it provides a smoother heat capacity curve. We have monitored the acceptance ratios for all $42$ streams and found values larger than $60\%$ for all simulations performed.  Thus, the overlap between neighboring temperatures is more than adequate. 

As previously mentioned, besides the acceptance ratios of swaps, we have also monitored individual acceptance ratios for the Metropolis sampling at each temperature. We have ensured that these acceptance ratios are between $40\%$ and $60\%$ by automatically adjusting the values of the maximal displacements for the path variables in the equilibration phase of the computation. Numerical experimentation has showed that in order to achieve a statistical error of less than $1k_B$ for heat capacities, it suffices to set the length of the data accumulation phase to $100$ blocks, for a total of $4$ million passes per temperature. The equilibration phase has consisted of $20$ blocks. We have therefore employed a number of data accumulation passes per temperature equal to the one utilized by Neirotti, Freeman, and Doll. This facilitates a direct comparison between the two heat capacities estimators introduced in the preceding section and the double virial estimator.

The discretization step $\epsilon_0$ entering the finite difference schemes has been set to $\epsilon_0 = 2^{-12}$. We mention that computer experimentation has shown that the numerical accuracy of the finite difference schemes is at least one thousand times smaller than the statistical error for all simulations performed and for a large range of $\epsilon_0$. Good values for $\epsilon_0$ are any inverse powers of two between $2^{-18}$ and $2^{-8}$.

We conclude this section by commenting on the evaluation of the errors involved in the determination of heat capacities. As opposed to energy estimators, heat capacity estimators are biased. This is apparent from Eq.~(\ref{eq:II2}).  In a general setting, let us denote by $X_i$ and $Y_i$ the block averages of two quantities $X$ and $Y$ and let us define 
\[ \bar{X}_n = \frac{1}{n}\sum_{i=1}^n X_i \quad \text{and} \quad \bar{Y}_n = \frac{1}{n}\sum_{i=1}^n Y_i.\]
Given a continuously differentiable function $f(x,y)$, we have
\[
f(\bar{X}_n, \bar{Y}_n) \to f( \left\langle X \right\rangle, \left\langle Y\right\rangle ) 
\]
almost surely, but unless $f(x,y)$ is linear in its variables, $f(\bar{X}_n, \bar{Y}_n)$ is a biased estimator of $f( \left\langle X \right\rangle, \left\langle Y\right\rangle ) $. In the limit that the variables $\bar{X}_n$ and $\bar{Y}_n$ have small fluctuations around their expected values, the following approximation holds:
\begin{eqnarray*}&&
\left\langle \left[f(\bar{X}_n, \bar{Y}_n) - f( \left\langle X \right\rangle, \left\langle Y\right\rangle ) \right]^2 \right\rangle \\ && \approx \bigg\langle \bigg[\frac{\partial}{\partial x} f(\left\langle X \right\rangle, \left\langle Y\right\rangle)(\bar{X}_n - \left\langle X \right\rangle) \\ && + \frac{\partial}{\partial y} f(\left\langle X \right\rangle, \left\langle Y\right\rangle)(\bar{Y}_n - \left\langle Y \right\rangle)  \bigg]^2 \bigg\rangle 
\end{eqnarray*}
As such, the mean square deviation for the quantity of interest is given by the variance of the quantity
\[\frac{\partial}{\partial x} f(\left\langle X \right\rangle, \left\langle Y\right\rangle)\bar{X}_n  \\ + \frac{\partial}{\partial y} f(\left\langle X \right\rangle, \left\langle Y\right\rangle)\bar{Y}_n, \]
variance that can be evaluated with the (again biased) estimator
\begin{eqnarray}
\label{eq:II31} \nonumber
\frac{1}{n(n-1)} \sum_{i=1}^n \left[\frac{\partial}{\partial x} f( \bar{X}_n, \bar{Y}_n)(X_i - \bar{X}_n)\right. \\ + \left. \frac{\partial}{\partial y} f(\bar{X}_n, \bar{Y}_n)(Y_i-\bar{Y}_n)\right]^2.
\end{eqnarray}
The error bars reported in the present work represent twice the square root of the above expression. For the heat capacity problem, $f(x,y) = x - y^2$ and the quantities $X_i$ and $Y_i$ represent block averages of the second order and the first order derivatives of the function $R_n(\mathbf{x},\bar{\mathbf{a}};\beta,\epsilon)$ around the point $\epsilon = 1$ [see Eqs.~(\ref{eq:II2}), (\ref{eq:II19}), and (\ref{eq:II20})].

\subsection{Numerical results}
\begin{figure}[!tbp] 
   \includegraphics[angle=0,width=7.0cm,clip=t]{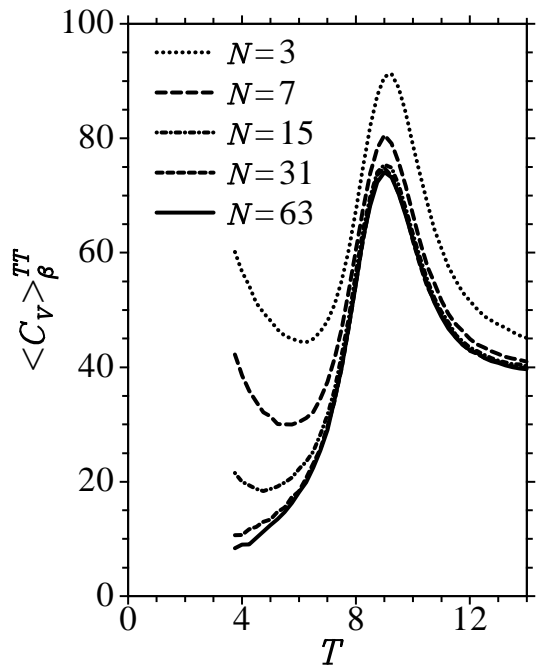} 
 \caption[sqr]
{\label{Fig:1}
Heat capacities (in units of $k_B$) computed with the TT-method estimator as a function of $T$ (in Kelvin) for several values of $N$. The error bars (two times the standard deviation) are comparable to the thickness of the drawing lines and are not plotted. 
}
\end{figure}

A graph of the heat capacity computed with the TT-method estimator as a function of temperature is found in Fig.~\ref{Fig:1} for each number of path variables employed. The sole exception is an additional run performed with a number of $N= 127$ path variables, which produces results virtually indistinguishable (i.e., the differences are smaller than the error bars) from the $N= 63$ results. Therefore, the remainder of the simulations have been performed using $N = 63$ path variables. Table~\ref{Tab:I} of Appendix~A contains the values obtained in the $N = 127$ simulation for $T= 4, 5, \ldots, 14$ as well as the associated error bars. We believe such values are useful both in the design of approximate quantum methods and as a numerical test for present and future path integral methods. 

\begin{figure}[!tbp] 
   \includegraphics[angle=0,width=7.0cm,clip=t]{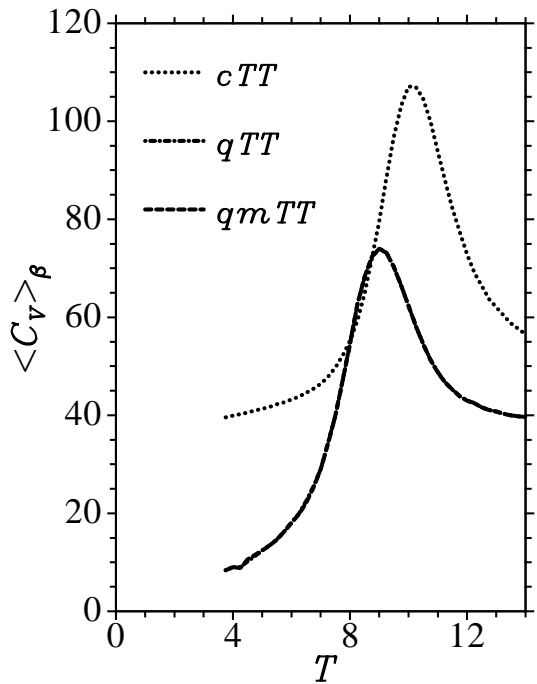} 
 \caption[sqr]
{\label{Fig:2}
Classical heat capacities ($cTT$) and quantum heat capacities by the TT-method estimator ($qTT$) and the modified TT-method estimator ($qmTT$)  as  functions of temperature. On this scale, the curves for the last two quantities overlap. The heat capacities are given in units of $k_B$, whereas the temperature is given in Kelvin. The number of path variables employed for the quantum results is $N = 63$. The error bars (two times the standard deviation) are comparable to the thickness of the drawing lines and are not plotted. 
}
\end{figure} 

\begin{figure}[!tbp] 
   \includegraphics[angle=270,width=8.0cm,clip=t]{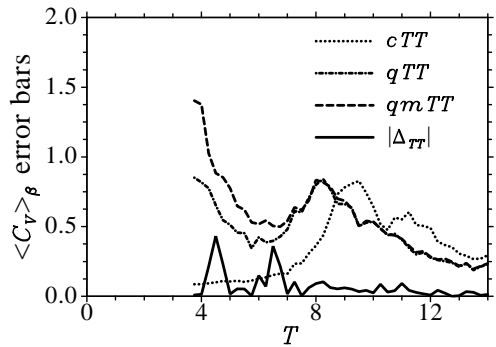} 
 \caption[sqr]
{\label{Fig:3}
Error bars (in units of $k_B$) for  classical heat capacities ($cTT$) and quantum heat capacities by the TT-method estimator ($qTT$) and the modified TT-method estimator ($qmTT$)  as  functions of $T$ (given in Kelvin). Also plotted (solid line) is the absolute value of the difference $\Delta_{TT} = \left\langle C_V \right\rangle^\text{TT}_\beta - \left\langle C_V \right\rangle^\text{mTT}_\beta$ between the heat capacity values computed with the help of the TT-method and modified TT-method estimators. 
}
\end{figure}

The modified TT-method estimator produces results similar to the direct TT-method estimator. As  shown in Fig.~\ref{Fig:2}, the curves for the two estimators are virtually indistinguishable. Fig.~\ref{Fig:2} also contains the classical heat capacity as a function of temperature. As apparent from Fig.~\ref{Fig:3}, the modified TT-method estimator has a larger variance in the low temperature region than the TT-method estimator. Though they seem to diverge to infinity as $T \to 0$, the error bars of both quantum estimators are comparable to the error bars of the classical estimator for the range of temperatures investigated. In the low temperature range, the error bars are about $10$ times smaller than those reported by Neirotti, Freeman, and Doll\cite{Nei00a} for the double virial estimator. Taking into consideration that the same number of Monte Carlo points has been employed,  the TT-method estimator is over $100$ times more efficient than the double virial estimator. We mention that the improvement has little to do with the path integral technique that has been utilized. Provided that enough path variables are considered, the variance of the estimators is independent of the path integral technique. At least in one other instance, such a significant improvement in the efficiency of a path integral technique has been eventually attributed to a superior estimator.\cite{Cha98, Dol99} 

\begin{figure}[!tbp] 
   \includegraphics[angle=0,width=7.0cm,clip=t]{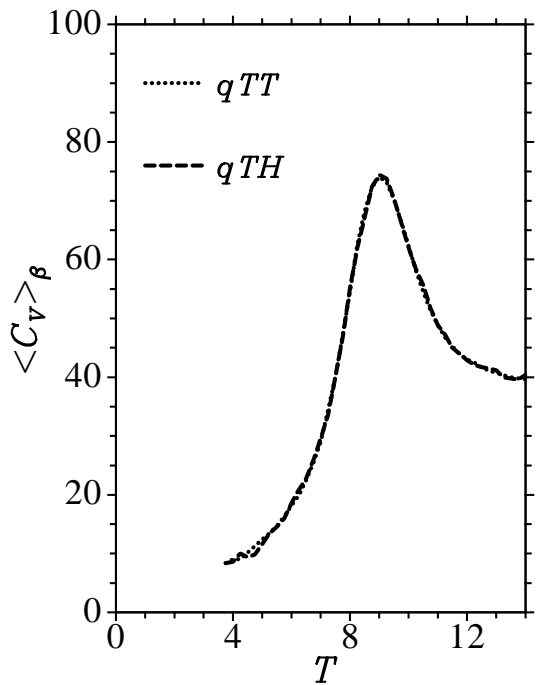} 
 \caption[sqr]
{\label{Fig:4}
Quantum heat capacities in units of $k_B$ by the TT-method  estimator ($qTT$) and by the TH-method estimator ($qTH$), respectively. The temperature is given in Kelvin. On this scale, the two curves overlap almost perfectly. The maximum difference between corresponding values on the curves is about $1.5 k_B$.
}
\end{figure}

As emphasized in Ref.~\onlinecite{Pre03d}, the agreement between the T-method and the H-method energy estimators constitutes an important test for the convergence of the path integral methods. The heat capacity analog is represented by the agreement between the TT-method and the TH-method estimators. The latter estimator is obtained by temperature differentiation of the H-method energy estimator. The temperature differentiation can be performed by finite difference in a way similar to the present implementation of the TT-method estimator. However, the evaluation of the H-method estimator requires knowledge of the first order derivatives of the potential. Since these derivatives have been computed anyway in the modified TT-method estimator simulation, we have also evaluated the H-method energy estimator in the respective simulation. A temperature differentiation with the help of the formula
\begin{equation}
\label{eq:II32}
\left\langle C_V\right\rangle^{TH}_{\beta_i} =-k_B \beta_i^2 \frac{\left\langle E\right\rangle^{H}_{\beta_{i+1}}-\left\langle E\right\rangle^{H}_{\beta_{i-1}}}{\beta_{i+1} - \beta_{i-1}} 
\end{equation}
has produced the curve in Fig.~\ref{Fig:4}, figure that also plots the TT-method heat capacity estimator, for comparison. The agreement between the two curves is surprisingly good. In fact, the maximum difference between the two curves is about $1.5 k_B$, a value that is comparable to the error bars achieved in the present simulations.  

We say that the agreement is surprisingly good because several factors concur against such an agreement. First, numerical differentiation of Monte Carlo data is in general a difficult task, unless the data at different temperatures are strongly correlated so that the resulting curve is smooth. In this regard, the parallel tempering technique is of great help because it brings  the necessary correlation into the simulation.  From the quality of the numerical differentiation, we estimate that the correlation is substantial. For instance, if the runs at different arbitrarily close temperatures are independent, the resulting curves fail to be continuous. If the correlation is of the type appearing in a Brownian motion, the resulting curves are continuous but not differentiable. In order for the curves to be differentiable, the correlation must be even stronger. Though such a strong correlation has been previously reported,\cite{Nei00b} we are not aware of any mathematical or numerical analysis attempting to quantify the strength of the parallel tempering correlation between averages at different temperatures. In the light of the application just presented, we believe such an analysis would be well justified.    

Second, the temperature step in the numerical differentiation is significantly larger than the step we have employed for the TT-method estimator. Fortunately, the quantum effects are strong and the dependence of the ensemble energy with the temperature is smooth. As a consequence, the accuracy of the finite-difference scheme is comparable to the statistical errors. 

 A third factor that could prevent an agreement between the TT-method and the TH-method heat capacity estimators is the lack of convergence of the path integral method employed. The agreement provides additional evidence that $N=63$ path variables are sufficient for the range of temperature studied and for the path integral technique utilized. 

Yet a fourth reason for disagreement is poor Monte Carlo sampling. Energy estimators are unbiased estimators, as opposed to heat capacity estimators, which are biased. As a consequence,  energy estimators and  heat capacity estimators generally have different sensitivities to the quality of the sampling, with the latter ones being more sensitive to quasiergodicity problems. This may result in  disagreement between the heat capacities computed with the help of  estimators and the ones computed by using energy differences of the type given by Eq.~(\ref{eq:II32}).

We conclude this section by noticing that the high-temperature part of the quantum heat capacity plotted in Fig.~\ref{Fig:2} does not coincide with the results reported in Ref.~\onlinecite{Nei00a}.  The cause of this difference is the fact that Neirotti, Freeman, and Doll have mistakenly utilized a confining potential with a radius $R_c = 4\sigma_{LJ}$ instead of $2\sigma_{LJ}$, the value they have reported.

\section{Summary and conclusions} 

The main result of the present paper is the finding that the evaluation of the main thermodynamic properties of a quantum canonical system, namely average energy and heat capacity, can be performed in a fast and reliable fashion without calls to first or second derivatives of the potential. This can be accomplished by a finite-difference scheme applied to the T-method energy estimator and TT-method heat capacity estimator, respectively. The derivation of these estimators is  rather trivial, consisting of simple temperature differentiations of the partition function. As emphasized in the introduction, the key observation is that the Feynman-Kac formula and its finite-dimensional approximations must be written in a form with the temperature dependence of the paths buried into the potential. Such a transformation is possible for all path integral techniques and it should constitute the starting point for the derivation of various energy and heat capacity estimators. 

We have also proposed an analytical heat capacity estimator, called the modified T-method estimator, that might prove useful whenever the first derivatives of the potential are available. However, this estimator has a slightly worse behavior at low temperature than the direct TT-method estimator and may be quite expensive for large-dimensional systems because of the quadratic scaling of the number of path integrals that must be computed with the dimensionality of the system. For example, in the case of the $\text{Ne}_{13}$ cluster, the code based on the modified heat capacity estimator has been $50\%$ slower than the code utilizing the finite-difference scheme.  

The heat capacity estimators utilized in the present paper have favorable variances when compared to the double virial estimators. This has been clearly demonstrated for a Lennard-Jones realization of $\text{Ne}_{13}$, a  realistic physical system that is representative of many other applications.   To the authors' knowledge, the heat capacities results obtained for the $\text{Ne}_{13}$ cluster are the most accurate to date. 

\begin{acknowledgments} The authors acknowledge support from the National
Science Foundation through awards No. CHE-0095053 and CHE-0131114. They
also wish to thank the Center for Advanced Scientific Computing and
Visualization (TCASCV) at Brown University for valuable assistance with
respect to the numerical simulations described in the present paper.
\end{acknowledgments}

\appendix
\section{Table of heat capacities}

\begingroup
\begin{table*}[!bthp]
\caption{
\label{Tab:I}
Heat capacities and error bars of the $\text{Ne}_{13}$ cluster as functions of temperature. A number of $N=127$ path variables have been utilized. The error bars are two times the standard deviation. The temperature is measured in Kelvin, whereas the heat capacities are given in units of $k_B$. The  heat capacity pick value, as obtained by maximizing a cubic spline  function interpolating the computed data, is $74.47 \pm 0.54~k_B$ and is attained at the temperature of $T_\text{peak} = 8.97~\text{K}$.}
\begin{tabular}{|c |c |c |c |c | c | c | c |}
\hline
$T$ & 4 & 5 & 6 & 7 & 8 & 9  \\ 
\hline 
$\left\langle C_V\right\rangle_\beta$& 8.26 $\pm$  0.80 &11.81 $\pm$ 0.52& 18.14 $\pm$ 0.44 &29.30$\pm$ 0.56& 54.74 $\pm$  0.86& 74.45 $\pm$   0.54 \\
\hline
\hline
$T$ &  10 & 11 & 12 & 13 & 14 & \\ 
\hline 
$\left\langle C_V\right\rangle_\beta$ &61.70$\pm$  0.51 & 48.87$\pm$  0.36 &  43.05 $\pm$  0.30 & 40.78 $\pm$ 0.23 & 40.09$\pm$ 0.27&  \\
\hline
\end{tabular}
\end{table*}
\endgroup



\end{document}